\documentclass[letterpaper, 10 pt, conference]{ieeeconf}  

\IEEEoverridecommandlockouts                              
\usepackage{amsmath, nccmath}
\usepackage[numbers]{natbib}
\usepackage{graphicx}
\usepackage{amsfonts}

\usepackage{bbm}

\newcommand{\pp}[1]{\left( #1 \right)}
\newcommand{\mb}{\mathbf}

\newcommand{\mc}{\mathcal}

\newcommand{\norm}[1]{\left|\left| #1 \right|\right|}
\newcommand{\expect}[2]{\mathbbm{E}_{#1}\left[ #2 \right]}

\title{\LARGE \bf
Two equalities expressing the determinant of a matrix in terms of expectations over matrix-vector products}
\author{Jascha Sohl-Dickstein
\\
\tt\small jaschasd@google.com}

\begin{document}

\maketitle
\thispagestyle{empty}
\pagestyle{empty}

\begin{abstract}
We introduce two equations expressing the inverse determinant of a full rank matrix $\mb A \in \mathbb R^{n \times n}$ in terms of expectations over matrix-vector products. The first relationship is $\left|\operatorname{det}\pp{\mb A}\right|^{-1} = \mathbb E_{\mb s \sim \mathbb \mathcal S^{n-1}}\left[\, \norm{ \mb A \mb s }^{-n} \right]$, where expectations are over vectors drawn uniformly on the surface of an $n$-dimensional radius one hypersphere. The second relationship is $\left|\operatorname{det}\pp{\mb A}\right|^{-1}  = \expect{\mb x \sim q}{\,p\pp{\mb A \mb x} /\, q\pp{\mb x}}$, where $p$ and $q$ are smooth distributions, and $q$ has full support.
\end{abstract}


\section{Derivation}
\label{sec derive}

\subsection{General relationship: $\left|\mb A\right|^{-1}  = \expect{\mb x \sim q}{p\pp{\mb A \mb x} /\, q\pp{\mb x}}$}
\label{sec general}
Let $\mb A\in \mathbb R^{n\times n}$ be a full rank matrix, $\left|\mb A\right|$ be the absolute determinant of $\mb A$, and $p$ and $q$ be smooth distributions over $\mathbb R^n$, where $q$ has full support. 
The inverse determinant can be related to an expectation over functions of matrix-vector products:
\begin{align}
  \left| \mb A \right|^{-1}
    &= \left| \mb A \right|^{-1} \int d^n\mb x\  p\pp{\mb x} \nonumber \\
    &= \int d^n\mb x\  p\pp{\mb A \mb x} \nonumber \\
    &= \int d^n\mb x\  q\pp{\mb x} \frac{p\pp{\mb A \mb x}}{q\pp{\mb x}} \nonumber \\
    &= \expect{\mb x\sim q}{\frac{p\pp{\mb A \mb x}}{q\pp{\mb x}}}
    \label{eq p q equality}
\end{align}

\subsection{Refined relationship: $\left|\mb A\right|^{-1} = \mathbb E_{\mb s \sim \mathbb \mathcal S^{n-1}}\left[ \norm{ \mb A \mb s }^{-n} \right]$}
\begin{figure}
\centering
\includegraphics[width=0.7\linewidth]{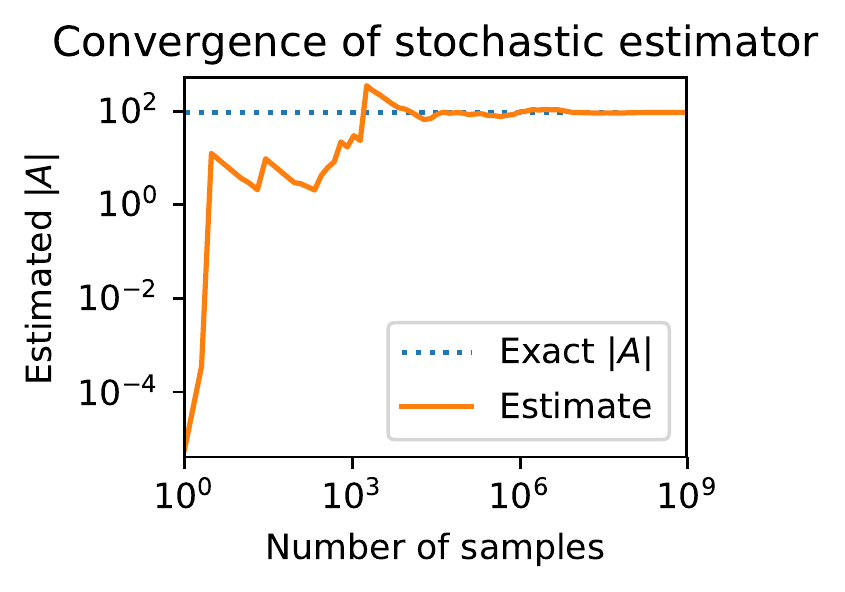}
\caption{
\textbf{An experimental validation of the relationship in Equation \ref{eq unbiased}.} 
We stochastically estimate the determinant of a matrix $\mb A$, by averaging $\norm{\mb A^{-1} \mb s}^{-n}$ over random unit-norm vectors $\mb s$. 
Here, $\mb A$ is a $10\times 10$ matrix, with iid, variance one, Gaussian entries.
\label{fig viz}}
\vspace{-2mm}
\end{figure}

Set $q = p = \mc N$, where $\mc N\pp{\mb x}$ is the probability density at $\mb x$ of a Gaussian with mean 0 and identity covariance. Additionally, let
$\expect{\mb s \sim S^{n-1}}{\cdot}$ be an expectation over vectors drawn uniformly on the surface of an $n$-dimensional radius one hypersphere, and $\expect{r \sim \chi\left(n\right)}{\cdot}$ be an expectation over a chi distribution with $n$ degrees of freedom. The following more refined relationship then follows:
\begin{align}
  \left| \mb A \right|^{-1}
    &= \expect{\mb x\sim \mc N}{\frac{\mc N\pp{\mb A \mb x}}{\mc N\pp{\mb x}}}\nonumber \\
    &= \expect{\mb x\sim \mc N}{\exp\left(
            \frac{1}{2}\left[
                \norm{\mb x}^2 - \norm{\mb A \mb x}^2
            \right]
        \right)} \nonumber  \\
    &= \expect{\mb s \sim S^{n-1}}{
      \expect{r \sim \chi\left(n\right)}{
        \exp\left(
            \frac{1}{2}\left[
                \norm{\mb s r}^2 - \norm{\mb A \mb s r}^2
            \right]
        \right)}} \nonumber  \\
    &= \expect{\mb s \sim S^{n-1}}{
      \expect{r \sim \chi\left(n\right)}{
        \exp\left(
            \frac{r^2}{2}\left[
                1 - \norm{\mb A \mb s}^2
            \right]
        \right)}} \nonumber  \\
    &= \expect{\mb s \sim S^{n-1}}{
            \norm{\mb A \mb s}^{-n}
        }
        \label{eq refined}
\end{align}
Because $\left|\mb A\right|^{-1} = \left|\mb A^{-1}\right|$, Equation \ref{eq refined} also provides an unbiased stochastic estimator for the determinant of a matrix, in terms of matrix-vector products with its inverse:
\begin{align}
    \left|\mb A\right| &= \expect{\mb s \sim S^{n-1}}{
            \norm{\mb A^{-1}\, \mb s}^{-n}
        }
        \label{eq unbiased}
\end{align}
Experimental validation of this relationship is presented in Figure \ref{fig viz}.

\section{Related Work}

The equality in Section \ref{sec general} has been used in physics \citep{zwanzig1954high}, but appears not to have been previously published as an explicit identity. 
Related expressions appear in work on ratios of moments of quadratic forms \citep{rukhin2009identities}, and in techniques for rewriting certain determinants in terms of integrals which can be evaluated by Monte Carlo \citep{weingarten1981monte,fucito1980proposal,finkenrath2018stochastic}.
In the special case of positive symmetric definite $\mb A$, Gaussian quadrature techniques have been used to stochastically estimate determinants \citep{bai1996some}. 
Other work derives
stochastic estimators of classes of log determinants \citep{han2015large,saibaba2017randomized,chen2019residual}. 
Hadamard's inequality can be reinterpreted as a stochastic upper bound on $\left| \mb A \right|$ in terms of the norms of row or column vectors \citep{hadamard1893resolution}.

\section{Discussion}
We hope that the stochastic estimators presented in this note will enable new Monte Carlo techniques for estimating, or stochastically bounding, functions of matrix determinants.
These relationships 
may be especially useful in machine learning for training and evaluating both normalizing flow models \citep{dinh2016density,rezende2015variational,karami2019invertible} and Gaussian process kernels \citep{rasmussen2003gaussian}.

\small

\section*{Acknowledgments}

Thank you to 
Alex Alemi, 
Anudhyan Boral,
Ricky Chen, 
Arnaud Doucet,
Guy Gur-Ari, 
Albin Jones,
Abhishek Kumar, 
Peyman Milanfar,
Jeffrey Pennington, 
Christian Szegedy, 
Srinivas Vasudevan,
and
Max Vladymyrov
for helpful discussion and links to related work.

\bibliographystyle{plainnat}
\bibliography{main}


\end{document}